\documentclass[letterpaper]{article}

\usepackage[T1]{fontenc}

\usepackage{geometry}
\usepackage{setspace}

\usepackage[style = chem-acs]{biblatex}
\usepackage{graphicx}
\usepackage{float}
\newfloat{scheme}{htbp}{los}
\floatname{scheme}{Scheme}
\floatname{chart}{Chart}
\newfloat{graph}{htbp}{loh}

\usepackage{chemformula} 
\usepackage[version = 4]{mhchem} 

\usepackage{authblk}
\author[1*]{Ze Zheng}
\author[2]{Olga Sergaeva}
\author[2,3]{Davide Rocco}
\author[2,3]{Luca Carletti}
\author[4]{Can Tao}
\author[4]{Yucheng Zhang}
\author[4]{Guoquan Zhang}
\author[2,3]{Costantino De Angelis}
\author[1]{Lei Xu}
\author[5]{Anatoly V. Zayats}
\author[1*]{Mohsen Rahmani}
\affil[1]{Advanced Optics and Photonics Laboratory, Department of Engineering, School of Science \& Technology, Nottingham Trent University, Nottingham NG11 8NS, UK}
\affil[2]{Department of Information Engineering, University of Brescia, via Branze 38, 25123 Brescia, Italy}
\affil[3]{National Institute of Optics, Consiglio Nazionale delle Ricerche, via Branze 45, 25123 Brescia, Italy}
\affil[4]{The MOE Key Laboratory of Weak Light Nonlinear Photonics, TEDA Applied Physics Institute and School of Physics, Nankai University, Tianjin 300457, China}
\affil[5]{Department of Physics and London Centre for Nanotechnology, King’s College London, Strand London WC2R 2LS, UK}

\title{Animated Ultraviolet Holography via Programmable Nonlinear Phase Transfer}
\date{*Email: ze.zheng@ntu.ac.uk;mohsen.rahmani@ntu.ac.uk}

\begin{document}

\maketitle

\begin{abstract}
Dynamic and programmable control of ultraviolet (UV) light remains challenging because of conventional high material absorption and the limited functionality of optics such as lenses, mirrors or filters, and optoelectronic devices, such as spatial phase modulators, over the UV range. Here, by employing the concept of virtual metasurfaces based on arrays of programmable optical patterns on a 600-nm-thick lithium niobate film as the nonlinear medium, we demonstrate nonlinear phase transfer, generating programmable holography in the UV range.  This arrangement provides a unique platform for generating animated holograms. A spatial light modulator encodes a computer-generated phase distribution onto a longer-wavelength pump beam, while second-harmonic generation transfers the programmed spatial phase to the UV field through the quadratic nonlinear interaction. Importantly, the ultrathin nonlinear film serves as a simple frequency-conversion interface, eliminating the need for nanostructure patterning or direct UV spatial modulation. By incorporating the nonlinear phase relationship into the hologram design, we experimentally reconstruct six distinct UV holographic patterns, with normalised cross-correlation values ranging from 0.68 to 0.83, demonstrating reliable UV holographic image reconstruction. As a result, the dynamic updating of the SLM enables animated UV holograms at 10 Hz, demonstrating reconfigurable UV holography using the same nonlinear film. The system also supports tunable SHG emission from 350 to 390 nm and dynamic beam steering at approximately 20 Hz. The results demonstrate the potential of virtual metasurfaces to realise software-defined UV holography by transferring spatial programming across optical frequencies. This opens new opportunities for reconfigurable short-wavelength UV beam shaping and structured illumination, without the need for UV-compatible optics or optoelectronics.


\end{abstract}

\section*{Keywords}

UV holography, flat optics, nonlinear optics, nonlinear holography

\section{Introduction}

UV light characterised by short wavelength and high photon energy offers unique capabilities that are difficult to access at visible and infrared wavelengths\cite{zhang2020low,schuster2021ultraviolet}. Its short wavelength enables high spatial resolution\cite{rocha2018experimental,mcfarlane2025advances}, while its high photon energy allows efficient excitation of electronic transitions\cite{schuster2021ultraviolet}, fluorescence\cite{lakowicz2006principles} and photochemical processes\cite{goti2024impact} in a wide range of materials. These properties have enabled a wide range of applications based on UV light, including photolithography\cite{ASML_Lithography_Principles,guan2024light,owa2023review}, photochemistry\cite{goti2024impact}, high-resolution imaging\cite{mcfarlane2025advances}, and spectroscopy\cite{lakowicz2006principles,rocha2018experimental}. There is an increasing demand for optical systems capable of flexibly and programmably controlling the spatial amplitude and phase of UV light. However, the same high photon energy that gives UV light its unique advantages also imposes serious restrictions on optical materials and modulation techniques\cite{zhang2020low,he2023ultra,ammenwerth2025dynamical}, as many materials exhibit strong absorption\cite{zhang2020low,he2023ultra}, photochemical degradation, or reduced lifetime\cite{moore2022fused} under UV illumination. These challenges become increasingly serious with shorter UV wavelengths, particularly in the deep UV range.

To improve such limitations, nonlinear metasurfaces can be exploited. Conventional metasurfaces are arrays of ultrathin, planar subwavelength resonators that can manipulate electromagnetic waves at subwavelength scales. The resonators, also called meta-atoms, can control wave phase, polarisation, and amplitude, enabling many cutting-edge applications such as ultra-thin lenses\cite{khorasaninejad2016metalenses}, quantum sources and quantum tomography\cite{wang2018quantum,zhang2022spatially}, as well as optical computing and image processing\cite{zhou2020flat}. Nonlinear metasurfaces can convert the incident light frequency to other frequencies\cite{vabishchevich2023nonlinear,rahmani2018nonlinear,gigli2022all,zheng2024broadband}. A key process for coherently changing the frequency of light is nonlinear conversion, which relies on strong light-matter interactions to exchange energy between frequencies efficiently. Traditionally, efficient optical frequency conversion required long propagation distances in bulk crystals to satisfy phase matching. Such bulk crystals do not meet the requirements of integrated photonics. The recent advent of metasurfaces, which facilitates strong field concentration\cite{shcherbakov2014enhanced}, has paved the way for efficient frequency conversion in subwavelength-size structures.

Initially, plasmonic (metallic) nanostructures were considered for harmonic generation\cite{li2015continuous}. However, metals absorb light in the visible and UV. Recently, high-refractive-index nanostructures, e.g., silicon and germanium, have shown improved performance due to reduced dissipative losses and the excitation of Mie-type resonances\cite{shcherbakov2014enhanced}. Despite their strong third-order nonlinearities, Si and Ge do not possess bulk second-order nonlinearities due to their centrosymmetric crystal structures. Thus, for the second-order nonlinear process (the strongest nonlinear interaction), III-V semiconductors have been explored for their high dielectric index and large second-order susceptibilities\cite{liu2016resonantly}. 
However, to date, there has been limited progress towards practical applications of nonlinear metasurfaces due to low conversion efficiency (ratio of output to input power).

The recently introduced virtual metasurfaces employ emulated optical patterns rather than physical meta-atoms that implement their functions either via arrays of engineered foci in free space\cite{zhu2023virtual} or via spatial light modulators (SLMs)\cite{zheng2026nonlinear}. SLMs can dynamically manipulate phase and amplitude in a two-dimensional pixel array, emulating physical metasurfaces\cite{neff1990two,savage2009digital}. The possibility of altering the shape of emulated patterns offers a unique opportunity to address the tunability limitations that remain a long-standing bottleneck for metasurfaces. 

Commercial SLM technologies are mainly based on liquid-crystal layers and micro-electro-mechanical systems (MEMS), enabling active and programmable control of the optical field, pixel by pixel\cite{yang2023review,allen2005micro,ho1998micro}. Although UV-compatible SLM technologies have been developed for selected wavelength ranges, their operating bandwidth, modulation capability, optical efficiency and power handling become increasingly restricted towards shorter UV wavelengths\cite{lin2004uv,lu1996ultraviolet,fong2010advances}. In particular, the availability of mature, high-resolution and dynamically programmable phase modulators remains limited in the deep-UV range (from 100~nm to 300~nm)\cite{won2025programmable}. As a result, achieving flexible and reconfigurable control of two-dimensional UV wavefronts remains significantly more challenging than in the visible and infrared domains.

An alternative approach is to avoid directly modulating the UV wavefront and instead encode the desired spatial information at a longer wavelength, where mature programmable modulators are available, before translating the encoded optical field into the UV through nonlinear frequency conversion\cite{won2025programmable,ackermann2023polarization}. In a nonlinear optical process, the interaction between an incident optical field and the nonlinear susceptibility of a material generates radiation at new frequencies through SHG, third-harmonic generation (THG), sum-frequency generation or four-wave mixing\cite{franken1961generation,boyd2008nonlinear}. Conventional bulk nonlinear systems can provide efficient frequency conversion, but the phase-matching conditions limit the spectral and angular acceptance, which can restrict the range of spatial frequencies that can be transferred\cite{boyd2008nonlinear,agrawal2000nonlinear,espinola2004raman,fejer1992quasi}.


\begin{figure}[htbp]
\centering
\fbox{\includegraphics[width=0.85\linewidth]{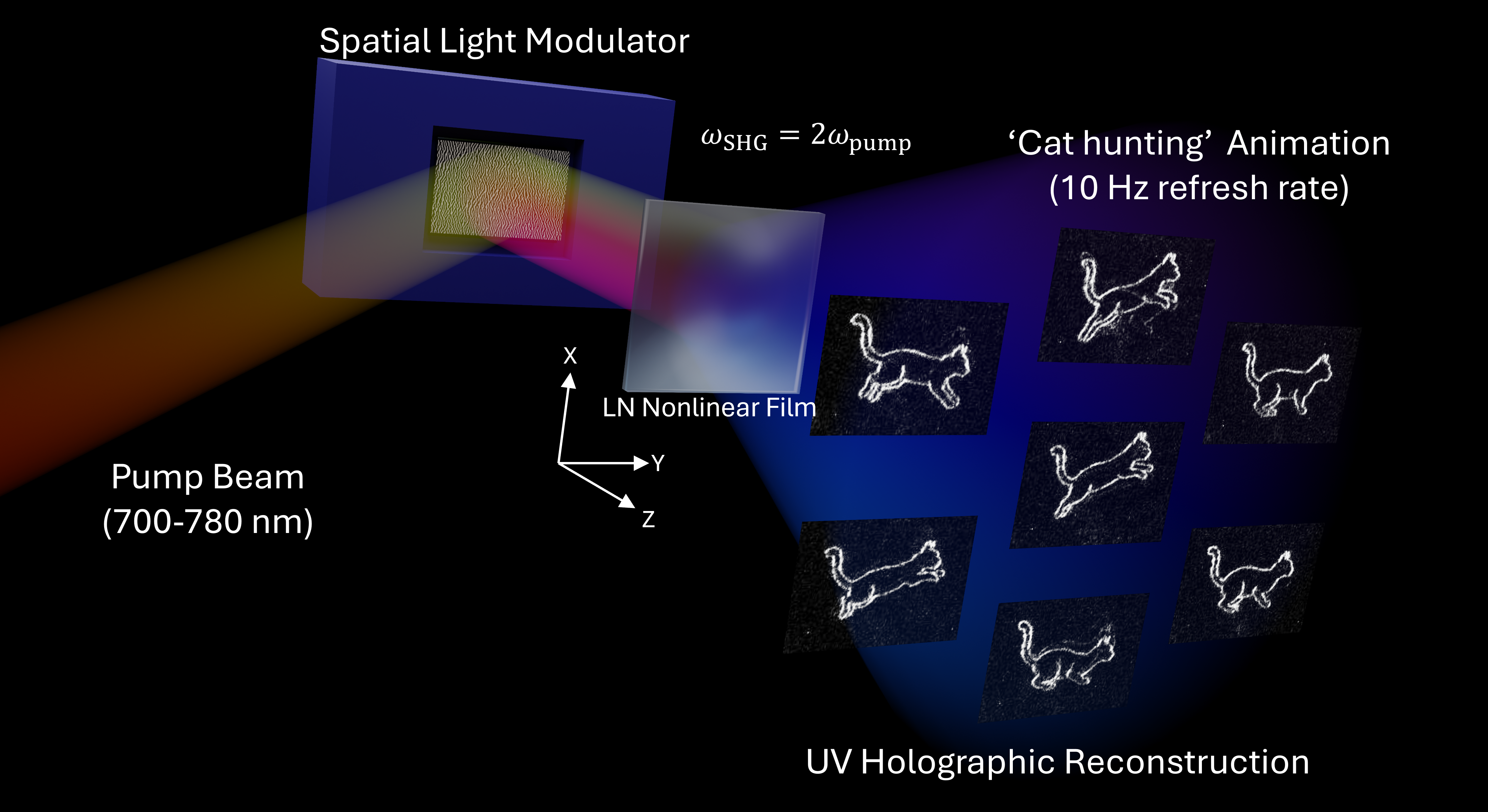}}
\caption{Schematic illustration of programmable ultraviolet holography based on nonlinear frequency conversion. A SLM encodes the desired phase distribution onto the fundamental pump beam. The modulated beam is incident on a nonlinear LN thin film, where SHG converts the optical field to the ultraviolet regime, with \(\omega_{\mathrm{SH}}=2\omega_{\mathrm{pump}}\). The encoded spatial phase information is transferred to the generated SH field, enabling programmable reconstruction of ultraviolet holographic patterns. By dynamically updating the phase holograms displayed on the SLM, time-varying UV holographic images, exemplified here by a “cat hunting” animation, can be generated at a 10 Hz refresh rate.}
\label{Figure:illu}
\end{figure}


Here, building upon the concept of virtual nonlinear metasurfaces, we demonstrate programmable ultraviolet holography using an unpatterned, 600-nm-thick lithium-niobate (LN) film combined with a SLM. Through nonlinear phase transfer in SHG, the spatial information encoded onto the fundamental wavefront is transferred to the ultraviolet field, enabling arbitrary holographic reconstruction and dynamic reconfiguration without requiring either direct UV modulation or nanostructure patterning. Using the same LN film, we experimentally demonstrate multiple UV holographic patterns and dynamically reconfigurable holographic animations at 10 and 20 Hz. In addition, tunable UV second-harmonic emission is achieved over the wavelength range of 350–390 nm. This approach provides a flexible route towards dynamically reconfigurable UV wavefront control and may offer opportunities for applications in UV imaging, lithography, spectroscopy and structured-light generation.

\section{Results and discussion}

\begin{figure}[htbp]
\centering
\fbox{\includegraphics[width=\linewidth]{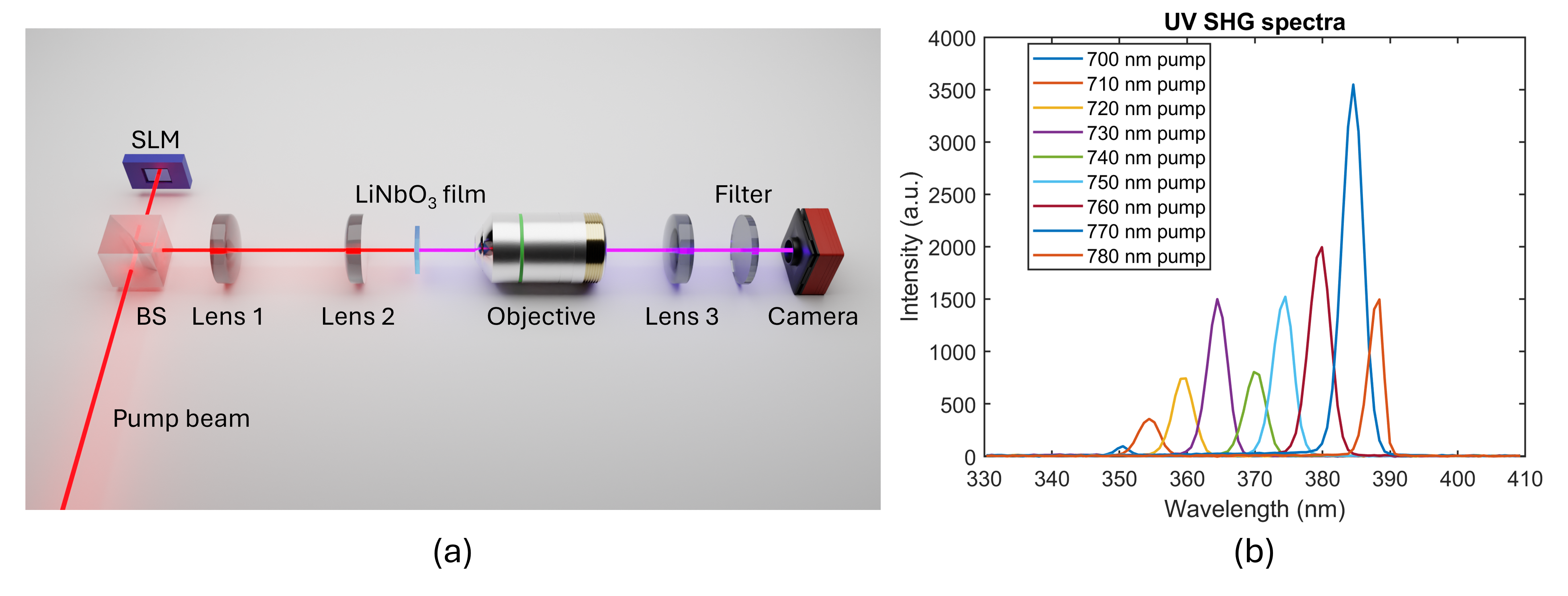}}
\caption{(a) Schematic of the experimental setup. (b) The measured spectra of the SHG emission from lithium niobate film in the UV range, tuning the pump wavelength from 700 to 780~nm.}
\label{Figure1}
\end{figure}

We use a 600-nm-thick LN film as the nonlinear platform, which has strong second-order nonlinearity. The generated SHG optical field can be estimated by treating the LN film as a sheet of second-harmonic (SH) polarisation, as shown below\cite{boyd2008nonlinear,novotny2012}:
\begin{equation}
\label{eq:eq1}
P_i^{(2)}(2\omega) = 2\,\varepsilon_0 \sum_{j,k} d_{ijk}\, E_j{(\omega)} E_k{(\omega)} \, ,
\end{equation}
where $\varepsilon_0$ is the vacuum permittivity, $P_i^{(2)}$ is the $i$-th Cartesian component of the SHG polarisation, $E_j{(\omega)}$ and $E_k{(\omega)}$ are the Cartesian components of the incident electric fields, 
and $d_{ijk}$ are the elements of the second-order nonlinear tensor. The second-order nonlinear tensor of the x-cut LN film used in this study can be found in Supporting Information section 1. The $x$- and $y$-directions lie in the film plane, and the z-direction is normal to the lithium niobate film. The crystallographic axis is aligned to the x-direction, while the incident polarisation is x-polarised (only $E_x(\omega)\neq0$), which can simplify the SH polarisation expressed as:
\begin{equation}
\label{eq:eq2}
P_x^{(2)}(2\omega) = 2\,\varepsilon_0d_{33}\, E_x^2{(x,y)} \, .
\end{equation}

\begin{figure}[htbp]
\centering
\fbox{\includegraphics[width=0.7\linewidth]{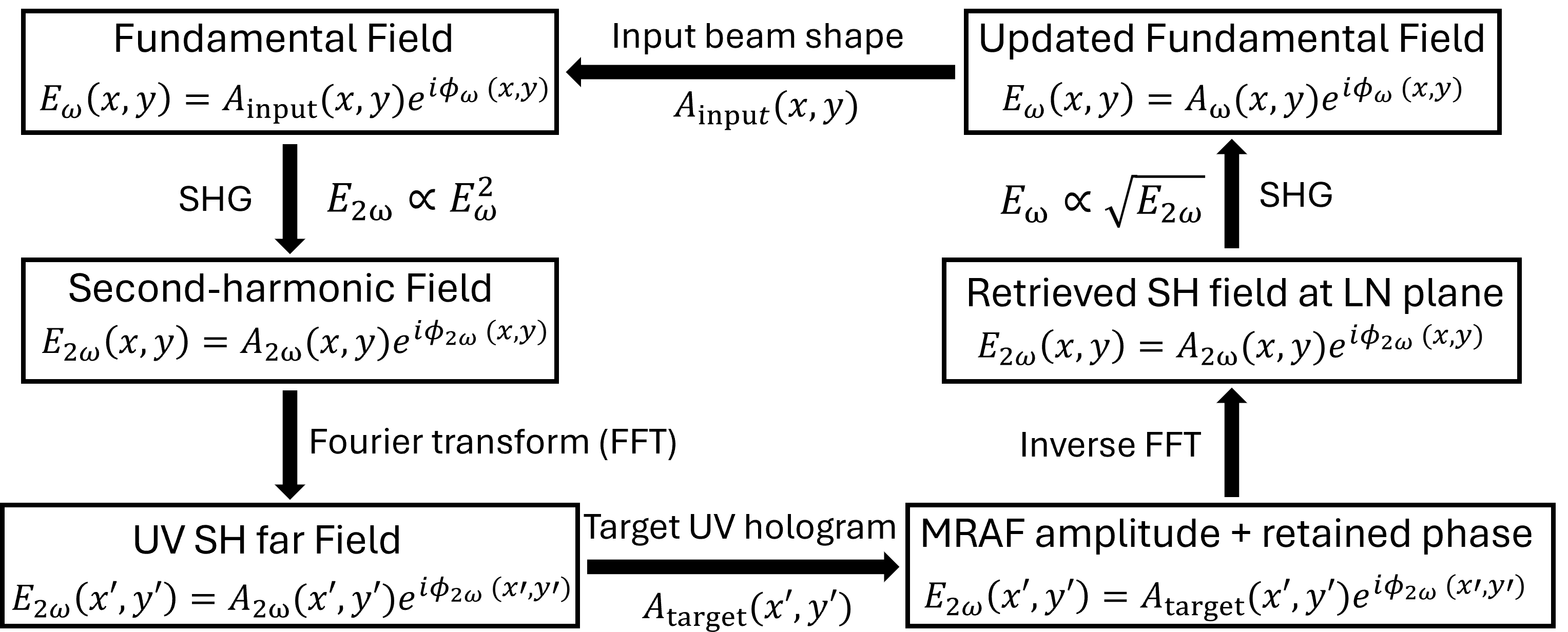}}
\caption{Schematic of the nonlinear Gerchberg–Saxton (GS) algorithm for UV SH hologram generation.}
\label{Figure2}
\end{figure}

The input spatial electric field distribution is modulated with a SLM, applying a phase pattern $\phi_\mathrm{SLM}(x,y)$ to the electric field. The amplitude distribution of the input electric field $A(x,y)$ would usually be Gaussian. Thus, the SH polarisation obtained with the input electric field ($E_x{(x,y)}=A(x,y)e^{i\phi_\mathrm{SLM}(x,y)}$) is
\begin{equation}
\label{eq:eq3}
P_x^{(2)}(2\omega) = 2\,\varepsilon_0d_{33}\, A^2(x,y)e^{i2\phi_\mathrm{SLM}(x,y)} \, .
\end{equation}
The key implication of Eq.~\ref{eq:eq3} is that the spatial phase of the generated SH field can be programmed entirely through the incident fundamental wavefront, without requiring any spatial patterning of the nonlinear susceptibility. At the nonlinear source plane, the SH phase follows twice the fundamental phase, while its amplitude scales quadratically with the incident field. Therefore, the unpatterned LN film acts as a nonlinear interface that transfers spatial phase information from the SLM-controlled fundamental beam into the UV. By incorporating this nonlinear field relationship into the hologram design, arbitrary target intensity patterns can be reconstructed at the SH frequency using the same nonlinear medium.

We next construct an optical system to experimentally validate the nonlinear phase-transfer mechanism and demonstrate programmable UV holography, as schematically illustrated in Fig. \ref{Figure1}(a). A tunable femtosecond laser with a pulse duration of approximately 150~fs and a repetition rate of 80~MHz, ranging from 700 to 780~nm, is used as the pump beam. The incident beam is directed by a beam splitter (BS) onto a reflective spatial light modulator (SLM), where the desired spatial phase distribution is encoded onto the fundamental wavefront. After reflection from the SLM, the modulated beam is redirected by the same BS into a 4-f imaging system consisting of Lens 1 and Lens 2. The 4-f system images the SLM plane onto the 600-nm thick LN film, thereby transferring the programmed spatial phase distribution of the fundamental field to the nonlinear interaction region. SHG in the LN film converts the modulated fundamental field into the UV, with the spatial phase modulation transferred to the generated SH field according to the nonlinear phase relationship described above. By tuning the fundamental wavelength from 700 to 780~nm, the generated SH wavelength can be continuously tuned from 350 to 390~nm (Fig. \ref{Figure1}(b)). The generated UV radiation is subsequently collected by an objective, and Lens 3 projects the far-field distribution of the SH field onto a UV-sensitive camera (CMOS camera), where the reconstructed nonlinear UV hologram is recorded.

To determine the spatial phase distribution required on the SLM for generating a desired UV hologram, we employ a modified nonlinear Gerchberg-Saxton (GS) algorithm, as schematically illustrated in Fig.~\ref{Figure2}. Unlike the conventional GS algorithm, the nonlinear relationship between the fundamental and SH fields is incorporated into each iteration. At the nonlinear source plane, the fundamental field is expressed as \(E_{\omega}(x,y)=A_{\omega}(x,y)\exp[i\phi_{\omega}(x,y)]\), where \(A_{\omega}(x,y)\) is the fixed incident amplitude distribution and \(\phi_{\omega}(x,y)\) is the phase to be encoded onto the SLM. The corresponding SH field is then calculated according to \(E_{2\omega}(x,y)\propto E_{\omega}^{2}(x,y)\), such that its amplitude and phase are related to those of the fundamental field through the second-order nonlinear interaction.

The calculated SH field is propagated to the hologram plane through a Fourier transform, where its amplitude distribution is constrained towards the desired target pattern while the phase is retained. A mixed-region amplitude freedom (MRAF) constraint is applied so that the target amplitude is mainly enforced within the signal region, while the field outside this region is allowed greater freedom to improve the reconstruction quality. The modified SH field is subsequently propagated back to the nonlinear source plane through an inverse Fourier transform. Owing to the phase relationship \(\phi_{2\omega}=2\phi_{\omega}\), the updated phase of the fundamental field is obtained by dividing the retrieved SH phase by two. This procedure is iteratively repeated until the calculated SH far-field intensity converges towards the desired UV holographic pattern. The resulting fundamental phase distribution is then encoded onto the SLM for the experimental reconstruction of the UV hologram.

\begin{figure}[htbp]
\centering
\fbox{\includegraphics[width=0.8\linewidth]{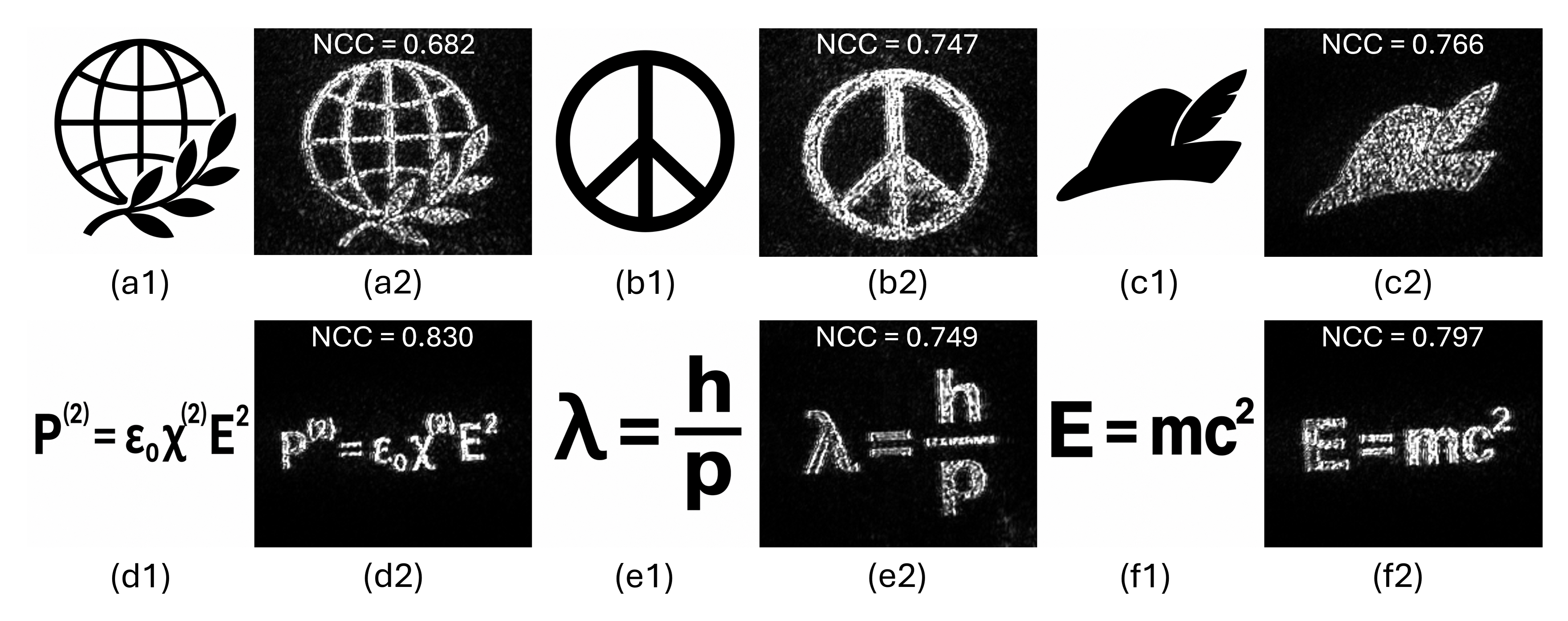}}
\caption{(a1-f1) Various designed target patterns and (a2-f2) corresponding experimentally reconstructed UV SH holographic images.}
\label{Figure3}
\end{figure}

We next employ the nonlinear GS algorithm together with the experimental optical system to demonstrate programmable UV holography. The designed target patterns are shown in Fig.~\ref{Figure3}(a1--f1), while the corresponding experimentally reconstructed UV SH holograms are presented in Fig.~\ref{Figure3}(a2--f2). The reconstructed holographic images show good agreement with the designed target patterns, demonstrating that the spatial information encoded onto the fundamental field can be successfully transferred to and reconstructed at the SH frequency at expected positions (see Supporting Information section 2). The reconstruction fidelity is further quantified using the normalised cross-correlation (NCC) between the target patterns and the corresponding experimental SH intensity distributions. NCC values of 0.68 to 0.83 are obtained across the six reconstructed patterns, with an average value of over 0.76, quantitatively supporting the observed agreement. The definition of NCC and detailed registration and calculation information can be found in the Supporting Information section 3.

\begin{figure}[htbp]
\centering
\fbox{\includegraphics[width=\linewidth]{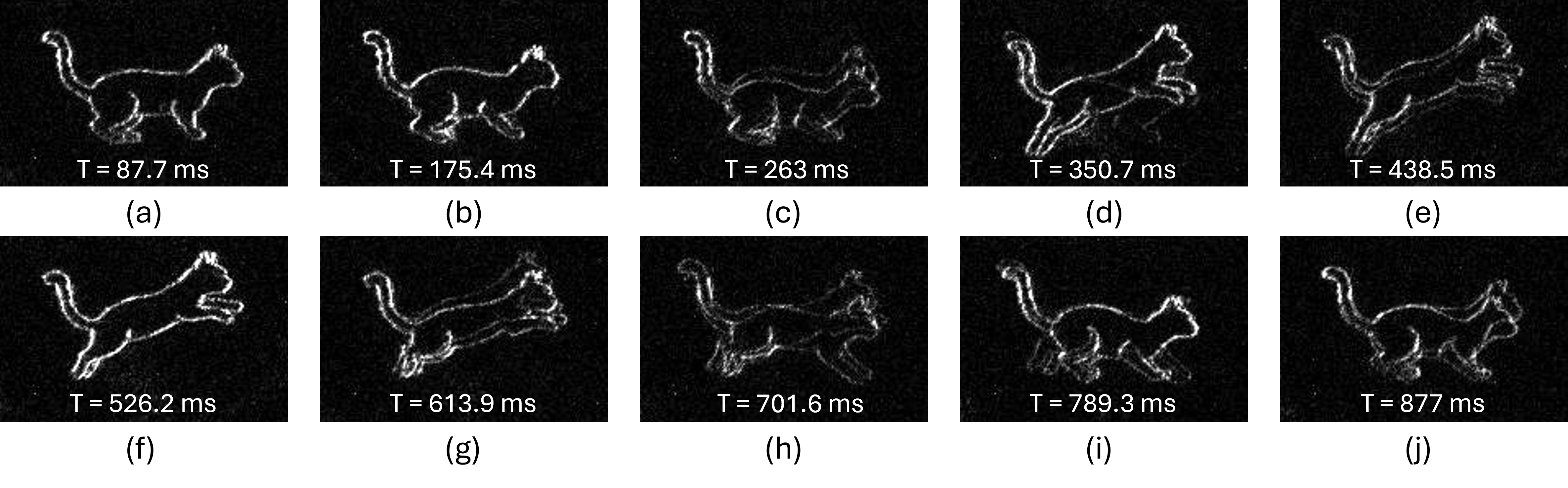}}
\caption{Video frames of the dynamically reconstructed UV SHG hologram. The SLM is operated at a modulation rate of 10 Hz, while the camera exposure time is 87.7 ms. See Supplementary Information for a full video.}
\label{Figure4}
\end{figure}

A central advantage of our approach is that the UV holographic wavefront can be dynamically reconfigured without modifying the nonlinear medium or introducing any active modulation element at the UV wavelength. Since the spatial information is entirely encoded in the fundamental field, updating the SLM phase pattern directly changes the reconstructed UV hologram. To demonstrate this capability, we calculate eight phase distributions corresponding to eight frames of a cat-hunting animation (see Supporting Information section 4) and sequentially display them on the SLM at a modulation rate of 10~Hz, corresponding to a refresh interval of 100~ms per holographic frame. The dynamically reconstructed UV holograms are recorded by the camera with an exposure time of 87.7~ms. The full cat-hunting animation can be found in Supplementary  Movie S1 and S2. Since the camera acquisition is not synchronised with the SLM refresh and operates at a different temporal sampling rate, ten camera frames are recorded over the eight-frame holographic pictures (Fig.~\ref{Figure4}). These results demonstrate dynamic reconfiguration of the nonlinear UV holographic wavefront through active modulation of the fundamental field. In fact, the frame rate of the holographic video is determined by the modulation rate of the SLM. An effective frame rate of approximately 20 Hz was achieved experimentally in our system, as verified from the recorded angular-scanning video in the Supporting Information section 5. The full angular-scanning video can be found in Supplementary Movie S3. These results demonstrate that the nonlinear frequency-conversion process preserves the reconfigurability provided by the fundamental-wavefront modulation, enabling dynamic UV holography without direct UV phase modulation.

\subsection{Conclusion}

We have demonstrated a programmable nonlinear UV holographic platform based on nonlinear virtual metasurfaces. We have exploited an unpatterned, 600-nm-thick LN film, as the nonlinear medium. Unlike approaches that rely on spatially engineered nonlinear nanostructures, our method separates holographic wavefront programming from the physical design of the nonlinear medium. Through the quadratic phase relationship in SHG, the spatial information encoded onto the fundamental beam is transferred to the UV, enabling multiple holographic functions using the same nonlinear interface. In addition to arbitrary static patterns, we demonstrate dynamic UV holographic animations at 10 and 20 Hz, confirming that the wavefront programmability of the fundamental beam can be retained following nonlinear frequency conversion. Together, these results establish a simple and reconfigurable approach to nonlinear UV holography without direct UV spatial modulation or nanostructure patterning, with potential applications in structured illumination, imaging and optical patterning.



\section*{Acknowledgements}

This research work has been kindly supported by the European Union (ERC UPIRI Grant No. 101170298). Views and opinions expressed are, however, those of the authors only and do not necessarily reflect those of the European Research Council. Neither the European Union nor the granting authority can be held responsible for them. Z.Z. and M.R. acknowledge the kind support from the UK Research and Innovation Future Leaders Fellowship (Grant No. MR/ Z000270/1). The authors appreciate the use of the NTU Imaging Suite, NTU High Performance Computing cluster Avicenna.

\section*{Supporting information}

Experimental procedures and additional characterisation data supporting the results of this study are provided in the Supporting Information. The captured videos of the UV holographic reconstruction are provided in the Supplementary Movie S1-3.

\printbibliography




  
  

\end{document}